\documentclass[12pt,a4paper,twoside]{article}

\author{Fabian Hassler}
\title{Topological Quantum Computing}
\date{Institute for Quantum Information\\* 
  RWTH Aachen University\\* Germany\\* 
\bigskip October 2024}

\usepackage[charter]{mathdesign}

\usepackage[top=12mm,bottom=12mm,left=25mm,right=25mm,head=12mm,includeheadfoot]{geometry}
\usepackage{tocloft}

\RequirePackage[nottoc,notlot,notlof]{tocbibind}

\usepackage{microtype} 

\usepackage{titlesec}
\titlespacing*{\section}{0pt}{1.8\baselineskip}{\baselineskip}

\usepackage[T1]{fontenc}

\usepackage{doi}

\usepackage{amsmath}

\usepackage{hyperref}
\hypersetup{colorlinks = true, urlcolor = blue, linkcolor = blue, citecolor = blue}

\usepackage{fancyhdr}

\usepackage{graphicx}

\usepackage{cite}

\usepackage[width=.90\textwidth]{caption}

\newcommand\pdag{{\phantom\dag}}

\begin{document}

\maketitle
\tableofcontents

\clearpage
\pagestyle{fancy}

\section{Introduction}

The concept of \emph{identical particles} is one of the most counterintuitive
features of many-body quantum mechanics, maybe second only to the concept of
entanglement. In fact, even the bosonic exchange statistics, which is the
closest to our classical world, has puzzled researchers time and time again.
The hallmark experiment by Hanbury Brown and Twiss \cite{hanbury:56}, showing
two-photon interference, was initially facing strong criticism: Brannen and
Fergusen, for example, stated that ``it would appear to the authors [\dots]
that if such a correlation did exist, it would call for a major revision of
some fundamental concepts in quantum mechanics''\cite{brannen:56}.  As a
reply, Purcell did point out that ``the electromagnetic field is a classical
field after all, which is why the Brown-Twiss effect only appears odd if one
looks at it from a particle point of view; its oddness being simply the
\emph{peculiarity of bosons}''\cite{silva:13}.  Going over to fermions, we do
not even  have a classical analog. The basic principle of fermions is that
they are described by anticommuting numbers. However, all observables, being
part of our everyday classical world, are bosonic. The consistency with
quantum mechanics then demands
`\emph{superselection}',\index{superselection} which means that all
observables have to be formed by an even number of fermionic operators.

The idea of using identical particles and their exchange statistics as a
resource for quantum computation is only a couple of decades old
\cite{freedman:02,kitaev:03}. The reason is that `simple' fermions and bosons,
which we are familiar with from the basic physics courses, are not useful for
this task.  Researchers have been looking at extensions of the concept of
exchange statistics of identical particles beyond the notion of fermions and
bosons. However, in three dimensions nothing interesting arises as all
possible parastatistics can be reduced to bosons and fermions.%
\footnote{The transformation of a general parastatistics to
bosons/fermions with potentially additional internal degrees of freedom is
called the Klein transformations.}

In two dimensions, we know that charged particles (with charge $q$)  pick up
the Aharonov-Bohm phase $q \Phi/\hbar$ when encircling a magnetic flux $\Phi$.
The Aharonov-Bohm phase is \emph{topological} in the sense that it does not
depend on the concrete trajectory taken by the particles but only on the
winding of the particle around the flux.  Composite (bosonic) particles
consisting of an electric charge $q$ and a magnetic flux $\Phi$ acquire a
phase $\theta =  q \Phi/2\hbar$ when exchanged.\footnote{The factor $\frac12$
arises as exchanging the particles only moves them halfway around each other.}
The value $\theta =0$ ($\theta= \pi$) corresponds to a composite boson
(fermion).  However, also other values are allowed which correspond to
Abelian anyons. In three dimensions, the construction does not work. In fact,
it is a theorem by Dirac \cite{dirac:31} that a consistent theory only allows
for pointlike magnetic charges  (called magnetic monopoles) that produce a
flux of size $\Phi_0 = n\times 2\pi\hbar/q$ leading to $\theta = n \pi$
($n\in\mathbb{Z}$). We can understand  Dirac's argument as follows: since
there is no well-defined winding number between two point particles in three
dimensions, there can be no topological phase which restricts the charge of
potential magnetic monopoles.

The general idea of how to use identical particles for topological quantum
computation is the following. In the standard gate model of quantum computers,
a calculation consists of three steps: (1) initialization in the state
$|i\rangle= |0,0,\dots\rangle$, (2) application of a gate $U$ (a general
unitary operator) on $|i\rangle$ which produces $|f\rangle = U|i\rangle$, (3)
measurement of the outcomes $o_j\in\{0,1\}$ in the computational basis with
probabilities $P(o_1,o_2, \dots) = |\langle o_1,o_2, \dots | f \rangle |^2$.
In a topological quantum computer, the three steps are replaced by operations
involving anyons \cite{preskill}: (1) the initialization is replaced by
splitting a pair of anyons out of the vacuum, (2) a gate is done by
\emph{braiding} the anyons with each other, (3) the measurement is performed
by fusion (pairwise annihilation), see Fig.~\ref{fig:fusion}($c$).

To obtain nontrivial gates, braiding should perform a unitary gate on the
ground state manifold. Since Abelian anyons only produce phases, they are not
useful for this task but rather non-Abelian anyons are needed. Note that the
computation on non-Abelian anyons is topological. The trajectories
(worldlines) in the process splitting-braiding-fusion form a knot in
space-time. The result of the computation does not depend on the concrete form
of the worldlines but only on the topology of the knot. It is in this respect
that the computation performed by non-Abelian anyons is robust.

The outline of the lecture is as follows: first, we will introduce the notion
of a topological superconductor. Then, we will show that Majorana zero modes
appear as zero energy solutions of the Bogoliubov-de Gennes equation
describing a spinless \emph{p}-wave superconductor in one dimension. The zero
modes will turn out to be  non-Abelian anyons called Ising anyons. However,
note that braiding of Ising anyons does not result in the application of any
arbitrary unitary gate.  Thus, we will introduce an alternative class of
anyons called `Fibonacci anyons'. We will show that they are in fact
universal, i.e., an arbitrary computation can be performed by braiding. There
are by now several reviews where further information on these subjects can be
found. For Majorana zero modes, see for example
Refs.~\cite{alicea:12,leijnse:12,beenakker:13}.  More information about
topological quantum computation can be found in
Refs.~\cite{preskill,pachos,nayak:08,stern:13,sarma:15}.

\section{Topological superconductors}\index{superconductor!topological}

In the context of quantum mechanics, topology refers to a static system
described by a Hamiltonian where the  properties of  certain eigenstates are
insensitive to small perturbations (disorder). In order that these states are
not affected by hybridization with other states of the system, the energy
necessarily has to be in an energy gap (bandgap) of the system such that
direct coupling is forbidden by energy conservation.\index{energy gap}

\subsection{Topological quantum number}\index{quantum number!topological}

A prime example of a topological system in one dimension (1D) is given by the
Jackiw-Rebbi (JR) model (1976) with the Hamiltonian
\cite{jackiw:76}\index{model!Jackiw-Rebbi}
\begin{equation}\label{eq:jr_model}
  H_\text{JR} =  v_F p \sigma^z + M(x) \sigma^x 
  = 
  \begin{pmatrix}
    -i \hbar v_F \partial_x &  M(x) \\
    M(x) & i \hbar v_F \partial_x
  \end{pmatrix},
\end{equation}
where $\sigma^j$ denote the Pauli matrices which represent an (artificial)
spin degree of freedom.  The first term describes the motion of a particle
with velocity $ \pm v_F$ and momentum $p = - i \hbar \partial_x$. The particle
is chiral as the direction `$\pm$' of motion depends on the spin. The second
term proportional to $M(x)$ leads to backscattering (as it couples the two
directions of motion) and the appearance of a `mass', i.e., a gap in the
energy spectrum, see below.

\begin{figure}
  \centering
  \includegraphics[width=\textwidth]{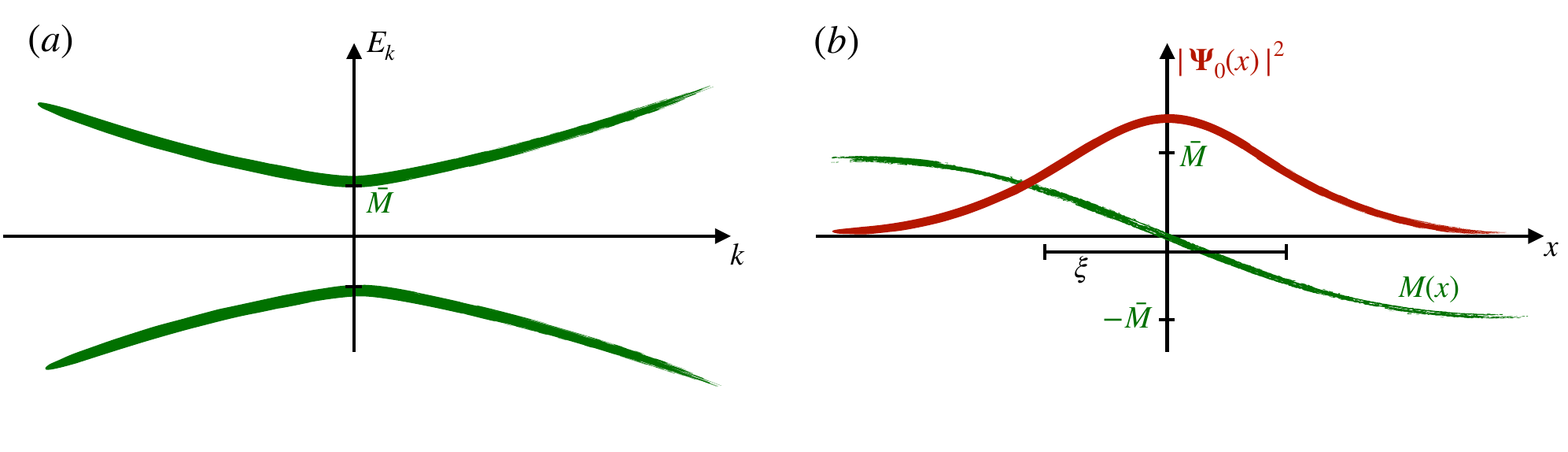}
  \caption{%
    ($a$) Spectrum of the Jackiw-Rebbi model. It consists of two bands, one at
    positive and one at negative energies $E_k$. For large momentum $\hbar k$,
    the spectrum assumes the form $E_k = \pm v_F \hbar |k|$. Close to
    $k\approx 0$, there is an anticrossing with a bandgap of size $2 \bar M$.
    ($b$)  Probability distribution $|\mathbf \Psi_0(x)|^2$ for finding the
    particle in the bound state with $E=0$ at the position $x$. It is peaked
    at the position $x\approx 0$ where the mass changes sign. The decay
    happens on the characteristic scale $\xi = \hbar v_F/\bar M$, which  is
    proportional to the inverse gap.
  }\label{fig:spekt_gap}
\end{figure}

Let us first discuss the case where the mass does not depend on position with
$M(x) \equiv \bar M$. In this case, the system is translation-invariant and we
can find eigenstates of the form
\begin{equation}\label{eq:eigenstates}
  \psi_k = e^{i k x} \mathbf v_k \,.
\end{equation}
Solving the eigenvalue equation $H_\text{JR} \psi_k = E_k \psi_k$ in this
case, leads to the result
\begin{equation}
  E_k = \pm \sqrt{(\hbar v_F k)^2 + \bar M^2}, \qquad \mathbf v_k = 
  \begin{pmatrix}
     E_k - k \\ \bar M
  \end{pmatrix}.
\end{equation}
Looking at the spectrum, cf.\ Fig.~\ref{fig:spekt_gap}($a$), we see that the
JR Hamiltonian is a model for a 1D semiconductor with a bandgap of $2 |\bar
M|$.  At first sight, it looks like the sign of $\bar M$ does not matter.

However, let us see what happens if we bring a semiconductor with a positive
bandgap $\bar M> 0$ in proximity to a semiconductor with a negative bandgap
$-\bar M$. We model this system by a mass $M(x)$ that depends on position and
assumes the asymptotic values $M(x) \to \pm \bar M$ for $x\to \mp\infty$, see
Fig.~\ref{fig:spekt_gap}($b$). We claim that, in this case, there will be  a
single state with energy $E=0$. We show this by directly computing the
eigenstate, i.e., solving the problem
\begin{equation}
  H_\text{JR} \mathbf \Psi_0 = -i \hbar v_F \sigma^z \partial_x\mathbf \Psi_0 
  + M(x) \sigma^x \mathbf \Psi_0  =0 \,.
\end{equation}
After a slight rearrangement, using $\sigma^z \sigma^x = i \sigma^y$, we
obtain
\begin{equation}
  \frac{d}{dx} \mathbf \Psi_0 (x) = \kappa(x) \sigma^y \mathbf \Psi_0(x), \qquad
  \text{with}\quad \kappa(x) = \frac{M(x)}{\hbar v_F}\,,
\end{equation}
with the solutions
\begin{equation}\label{eq:psi0}
  \mathbf \Psi_0 = \mathcal{N}\exp \left[ \pm \int_0^x\!dx'\,\kappa(x')\right]
  \boldsymbol
  \chiup_{\pm_y},\, \qquad  \boldsymbol \chiup_{\pm_y} = \frac{1}{\sqrt 2} \begin{pmatrix}
  e^{i\pi/4} \\ \pm e^{-i\pi/4} \end{pmatrix},
\end{equation}
where $\mathcal{N}>0$ is the normalization constant and $\boldsymbol \chiup_{\pm_y}$
denote the eigenvectors of $\sigma^y$ to the eigenvalues $\pm 1$.

For our choice of $M(x)$, we have that 
\begin{equation}
  \int_0^x\!dx'\,\kappa(x')  \to \mp \infty, \qquad x \to \pm \infty\,.
\end{equation}
In order for the state $\mathbf \Psi_0$ to be normalizable, we are only allowed to
take the `$+$' sign in Eq.~\eqref{eq:psi0}. Concluding, we have found the
bound eigenstate $ \mathbf \Psi_0(x)  \propto\exp \left[
\int_0^x\!dx'\,\kappa(x')\right] \boldsymbol \chiup_{+_y}$ of the JR model at energy
zero. The rest of the spectrum consists of extended states above the gap with
energy $|E| \geq \bar M$. The energy of the state $\Psi_0$ is within the gap.

Recapitulating the argument that has led to the identification of $\mathbf
\Psi_0$, we realize that the presence of the state is independent of the
concrete form of $M(x)$ and only depends on the sign of the asymptotic values
for $x\to \pm \infty$. It is in this sense that the state $\Psi_0$ is
topological and is insensitive to disorder. In fact, one calls
$\mathcal{Q}=\operatorname{sgn} M$ a ($\mathbb Z_2$) topological charge that
can take values of $\pm 1$. When bringing two systems with opposite
topological charges into proximity, a bound state within the gap is trapped at
the interface.  This state is insensitive to disorder because its presence is
guaranteed by the properties (topological charge) of the bulk away from the
interface.

\subsection{Kitaev model}

Kitaev (2001) introduced a Hamiltonian (called the Kitaev model) that
implements the JR model in a one-dimensional (1D) superconducting system. The
model is given by \cite{kitaev:01}\index{model!Kitaev}
\begin{equation}\label{eq:kitaev_model}
  H_\text{K} = \sum_p \xi_p c^\dag_p c_p + 
  \frac12 \sum_p \Delta\, p \,(c_{-p}c_p +
  c^\dag_p c^\dag_{-p}),
\end{equation}
where $c_p$  are fermionic annihilation operators that obey the canonical
anticommutation relation $\{c_p, c_{q}^\dag\} =\delta_{p,q}$, $\{c_p, c_q\} =
0$. The first term described spinless electrons with momentum $p$ and energy
$\xi_p = p^2/2m -\mu$ relative to the chemical potential $\mu$. The second
term proportional to $\Delta>0$ describes the superconducting
pairing.\footnote{The pairing is of $p$-type because there is no $s$-wave
pairing for spinless electrons.} As we will see below, the fact that the
problem is realized in a superconductor has crucial implications for the
properties of the bound state $\mathbf \Psi_0$. In fact, it turns out that the
resulting excitations at the interface are Majorana zero modes that follow
non-Abelian statistics useful for quantum computing.

In order to make contact with the JR model, we write the model in Nambu space
\begin{equation}
  H_\text{K} =\frac12 \sum_p \mathbf C_p^\dag H_\text{BdG}(p) \mathbf C_p
\end{equation}
with $\mathbf C_p = (c_p, c^\dag_{-p})^T$ and where we have introduced the
Bogoliubov-de Gennes Hamiltonian
\begin{equation}\label{eq:h_bdg}
  H_\text{BdG}(p) = \xi_p \tau^z + \Delta\, p \,\tau^x
\end{equation}
with $\tau^j$ the Pauli matrices acting on the Nambu space. Due to the fact
that the Hamiltonian $H_\text{BdG}$ is derived from $H_\text{K}$ by doubling
the degrees of freedom when going over from $c_p$ to $\mathbf C_p$, it enjoys the
relation $ \tau^x H_\text{BdG}(-p) \tau^x = - H_\text{BdG}(p)$. As a result,
each eigenmode of $H_\text{K}$ leads to two eigenstates: for each eigenstate
$\mathbf\Psi_E(p)$ of $H_\text{BdG}$ at energy $E$, there is a state $\tau^x \mathbf
\Psi_E(-p)$ at energy $-E$.

As above, we are interested in states close to $E=0$. The Hamiltonian
$H_\text{BdG}$ allows for such a state only when $p=0$ (such that the second
term vanishes) and $\mu=0$ (such that the first term vanishes). To describe
the physics near the band closing at $\mu =0$ and $p=0$, we expand around
$p=0$:\footnote{Treating the rest as a perturbation that can be included in
principle later on.}
\begin{equation}
  H_\text{BdG} \approx -\mu \tau^z + \Delta \,p\, \tau^x\,.
\end{equation}
We see that $H_\text{BdG}$ is, in fact, the JR model with the replacement
\begin{align}
  \sigma^z &\mapsto \tau^x,&  v_F \mapsto \Delta, \\
  \sigma^x &\mapsto -\tau^z, &  M \mapsto \mu\,.
\end{align}
With that, we can identify the sign of $\mu$ with the topological charge.
Indeed, the Kitaev model $H_\text{K}$ describes a topological superconductor
when $\mu >0$. For $\mu<0$, the electrons are depleted from the wire, and the
state is a conventional insulator.

\begin{figure}
  \centering
  \includegraphics[width=\textwidth]{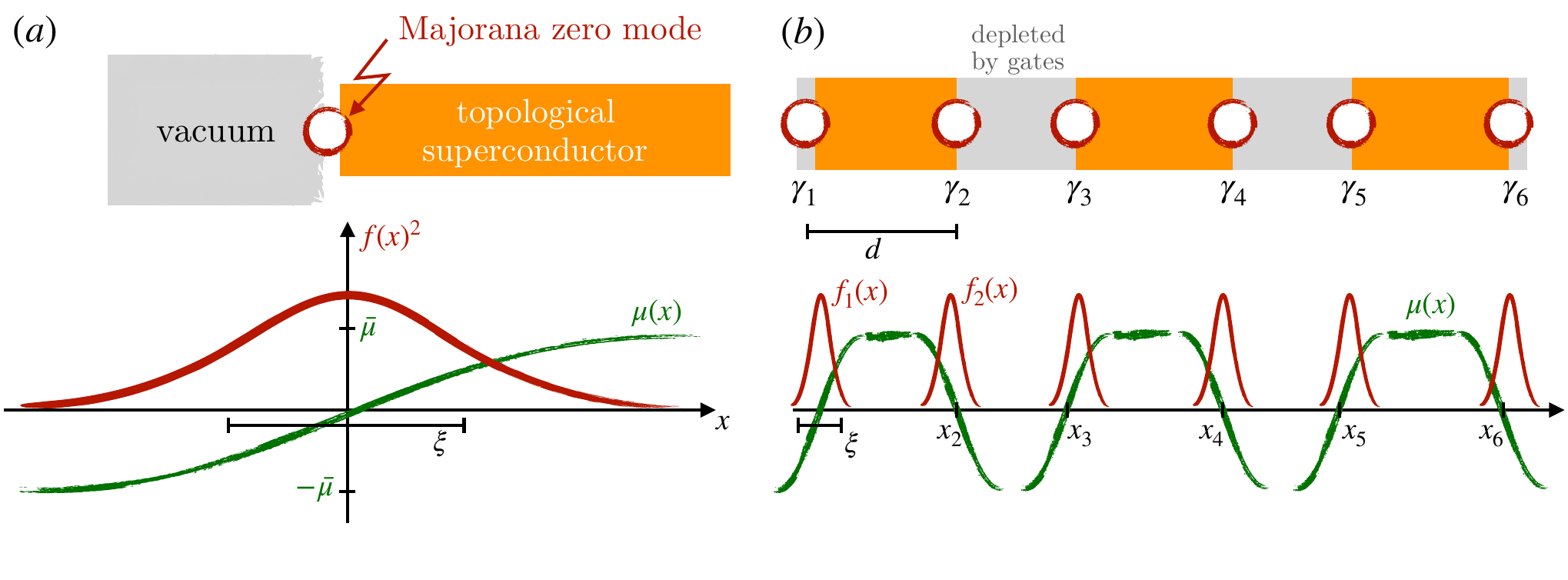}
  \caption{%
    ($a$) A Majorana zero mode appears at the interface between a 1D
    topological superconductor and the vacuum. The vacuum is modeled by a
    negative chemical potential $\mu$ such that there are no electrons
    present.  ($b$) Segments of a topological superconductor can be depleted
    by gates. In our effective model, the gates lead to a negative chemical
    potential. At each domain wall  (interface between the `vacuum' and the
    superconductor) at position $x_j$ a Majorana zero mode emerges. Due to the
    overlap proportional to $\exp(-d/\xi)$ with $\xi = \hbar \Delta/\bar \mu$,
    these modes are not exactly at zero energy but at an energy $\epsilon
    \simeq \bar \mu e^{-d/\xi}$.
  }\label{fig:mzm}
\end{figure}

The mapping of $H_\text{BdG}$ onto the JR-model allows us to predict that an
end state will appear at the interface of a topological superconductor ($\mu
>0$) to vacuum ($\mu<0$). In particular, we model the situation in
Fig.~\ref{fig:mzm} by a chemical potential $\mu(x)$ that changes sign at
$x=0$. The Hamiltonian $H_\text{BdG}$ has the bound state
\begin{equation}
  \mathbf \Psi_0(x) = f(x)\boldsymbol \chiup_{+_y}, \qquad f(x) = \mathcal{N} \exp\left[  
    \int_0^x \!dx'\, \mu(x') /\hbar\Delta 
  \right] 
\end{equation}
at energy $E=0$, which  is localized close to the interface at $x=0$. It is
separated by a gap $\bar \mu$  from the extended states.

\subsection{Majorana zero modes}

\index{Majorana zero mode}
As we have seen before, a single eigenmode of $H_\text{K}$ with energy $E>0$
corresponds to two states of $H_\text{BdG}$ with energy $\pm E$. It is thus
interesting to understand what the single state of $H_\text{BdG}$ at $E=0$
corresponds to. To this end, we calculate the second-quantized operator
\begin{equation}
  \gamma =  \sqrt 2\sum_p \int\!dx\,\mathbf \Psi_0(x)^* \cdot \mathbf C_p\, 
  e^{i p x/\hbar} 
  = \int\!dx\,f(x) \Bigl[ e^{i\pi/4} \psi(x) +
    e^{-i\pi/4}\psi^\dag(x)  \Bigr]
\end{equation}
that corresponds to the eigenstate $\mathbf \Psi_0$; here, $\psi(x) = \sum_p
e^{ipx/\hbar} c_p$ is the field operator that annihilates a particle at the
position $x$ and fulfills the anticommutation relations $\{\psi(x),
\psi^\dag(x')\} = \delta(x-x')$, $\{\psi(x), \psi(x')\} = 0$.

The fermionic mode $\gamma$ is special in that it is Hermitian with
$\gamma^\dag = \gamma$. In physical terms, this means that annihilating an
excitation in the mode $\gamma$ is the same as creating one. Such an
excitation is only possible at zero energy which corresponds to the chemical
potential of the superconductor. Due to the aforementioned Hermiticity, the
excitation described by $\gamma$ is called \emph{Majorana zero
mode}.\index{Majorana zero mode} With a straightforward calculation, one
can show that $\gamma^2 = \gamma^\dag \gamma =1$.  Generalizing these
properties to multiple points $x_j$ where the chemical potential $\mu$ changes
sign, we obtain a Majorana zero mode $\gamma_j$ at each interface (domain
wall) at $x_j$.  These modes fulfill the Clifford algebra
\begin{equation}
  \{ \gamma_i , \gamma_j \} = 2\delta_{i,j}
\end{equation}
which encodes the normalization $\gamma_j^2 =1$ together with the fermionic
statistics $\gamma_i \gamma_j = - \gamma_j \gamma_i$.

As we have seen before, a single Majorana zero mode $\gamma$ that is formed
at the position where the chemical potential changes sign is pinned at zero
energy. We can understand this in algebraic terms by noting that the effective
low energy Hamiltonian $H_\text{eff}$ can only involve the excitation $\gamma$
as the other excitations are at the energy $\bar \mu$. Due to
superselection\index{superselection}
(see below), each term of a Hamiltonian has to involve an even  power of
fermionic operators. However, all such terms are trivial as $\gamma^{2n} =
(\gamma^2)^n = 1$ so the effective Hamiltonian vanishes.

This argument does not work any more as soon as two Majorana zero modes
$\gamma_1$ and $\gamma_2$ are present. In fact, there is only one  Hermitian
combination $i \gamma_1 \gamma_2$ that involves an even number of fermionic
operators. So the effective Hamiltonian has to be of the form
\begin{equation}
   H_\text{eff} = \epsilon\, i \gamma_1 \gamma_2\,.
 \end{equation}
In fact, the energy $\epsilon$ is approximately given by the overlap
\begin{equation}
  \epsilon \approx \langle f_1 | H_\text{BdG} | f_2 \rangle  \simeq \bar\mu
  \int\!dx\,f(x-x_1)^* f(x-x_2) \simeq \bar \mu e^{-d/\xi}
\end{equation}
with $d =|x_2-x_1|$, the distance between the zero modes, and $\xi = \hbar
\Delta/ \bar \mu$, the decay length that is proportional to the inverse gap.

\section{Majorana qubits}

We have seen that the low energy properties of a topological superconductor
are given by the Majorana zero modes $\gamma_j$ that are pinned at the domain
walls where the sign of $\mu$ changes. We are interested in the Hilbert space
that can be accessed by acting with the modes $\gamma_j$ on the ground state
$|0\rangle$.  It is an easy exercise in algebra to show that given a set of
$2N$ Majorana zero modes, we can construct $N$ Dirac fermions $c_j$ via 
\begin{align}\label{eq:dirac}
  c_j &= \frac12 (\gamma_{2j-1} + i \gamma_{2j}), & 
  c^\dag_j & =\frac12 (\gamma_{2j-1} - i \gamma_{2j}).
\end{align}
The Dirac fermions fulfill  the canonical anticommutation relations
$\{c^\pdag_i,c^\dag_j\} = \delta_{i,j}$ and $\{c_i, c_j\}=0$.

The Hilbert space of a single fermionic mode ($N=1$) is two-dimensional: the
mode is either filled or empty, distinguished by the eigenvalue of the number
operators $n_j = c^\dag_j c^\pdag_j$ which have eigenvalues 0 or
1.\footnote{Note that $n_j$ is idempotent as $n_j^2= c^\dag_j c^\pdag_j
c^\dag_j c^\pdag_j = c^\dag_j (1-c^\dag_j c^\pdag_j ) c^\pdag_j = n_j $, which
proves that the eigenvalues of $n_j$ are 0 or 1.
} Operators that will turn out to be important in the following discussion
are the fermion parity operators $\mathcal{P}_j = 1- 2 n_j = (-1)^{n_j}$,
which have the eigenvalue $+1$ if the number of fermions is even and $-1$
if the number of fermions is odd. In terms of the Majorana operators,
the parity operators assume the simple form
\begin{equation}\label{eq:parity}
  \mathcal{P}_j = -i \gamma_{2j-1} \gamma_{2j}.
\end{equation}

If we think about an implementation for a quantum computer, we are used to the
example of a spin-$\tfrac12$ particle which is a model system for a generic
two-level system \cite{feynman:3}. However, we can ask ourselves the question
whether we can also use the many-body Fock space for quantum computation
purposes. We know that the occupation states $|n_1,n_2, \dots, n_N \rangle$
with $n_j\in\{0,1\}$ form a basis for the $N$-mode fermionic Fock space
generated by the creation operators $c^\dag_j$, $j\in\{1,\dots,N\}$, starting
from the vacuum state denoted by $|0\rangle$. The Fock space has dimension
$2^N$ (each mode can be either occupied or empty). Thus counting the degrees
of freedom, we are tempted to conclude that a fermionic system with $N$ modes
emulates $N$ qubits. In the next section, we will see that this naïve
counting argument is not completely correct as it violates the so-called
superselection\index{superselection} rule. 

\subsection{Fermionic quantum computation}\label{sec:fermion_qc}

Expressing a Hamiltonian $H$ or in fact any physical observable $A$ which are
bosonic operators in terms of fermionic creation and annihilation operators,
we are bound to only include terms where an even number of fermion operators
appear.\footnote{From the correspondence principle, we know that for large
quantum numbers the expectation values of operators for physical observables
should behave like (real) numbers. Due to the anticommutation relation of
fermionic operators, the correspondence principle for a potential fermionic
observable would instead lead to anticommuting Graßmann numbers on the
classical level.} The result is that the total fermion parity $\mathcal{P} =
\prod_{j} \mathcal{P}_j = (-1)^{\sum_j n_j}$ is strictly conserved in a closed
system; the reason for this is the fact that
\begin{equation}\label{eq:parity_on_hamilton}
  \mathcal{P} A \mathcal{P} =A
\end{equation}
which follows from $\mathcal{P} c_j \mathcal{P} = -c_j$ and the fact that each
term in $A$ involves an even number of fermionic operators.  Note that the
superconducting Hamiltonian \eqref{eq:kitaev_model} conserves the total
fermion parity even so the number of fermions is not conserved. Due to this
constraint, we have the following superselection rule: given two states in a
fermionic Fock space $|\psi_+\rangle$ and $|\psi_-\rangle$ with different
fermion parity, $\mathcal{P} |\psi_\pm\rangle = \pm |\psi_\pm \rangle$ we have
\begin{equation}\label{eq:superselection}
  \langle \psi_- | A | \psi_+ \rangle =\langle \psi_- | \mathcal{P} A
  \mathcal{P}| \psi_+ \rangle = - \langle \psi_- | A | \psi_+ \rangle =  0
\end{equation}
for all observables $A$.  Thus, there is no point in making superpositions
between states of different parity as there will be no effect on any
observable. We can restrict ourselves to one superselection sector and
keep the total fermion parity fixed with either $\mathcal{P} = +1$ or
$\mathcal{P}=-1$. The conclusion of this argument is that out of the $2^N$
states in a fermionic Fock space, only $2^{N-1}$ can be effectively used for
quantum computation purposes.

A further restriction to quantum computation using fermions arises from the
fact that noninteracting fermions subject to beam splitters, phase-shifters
(delay lines), measurements of the state of a single electron (so-called
fermionic linear optics) does not in fact lead to any entanglement
\cite{terhal:02}. In order to generate entanglement, we need to add parity
measurement of two electrons which effectively involves interactions between
different electrons \cite{beenakker:04b,hassler:13}.

\subsection{Encoding of a qubit}

\index{qubit!Majorana}
\begin{figure}[tb]
  \centering
  \includegraphics[width=.8\textwidth]{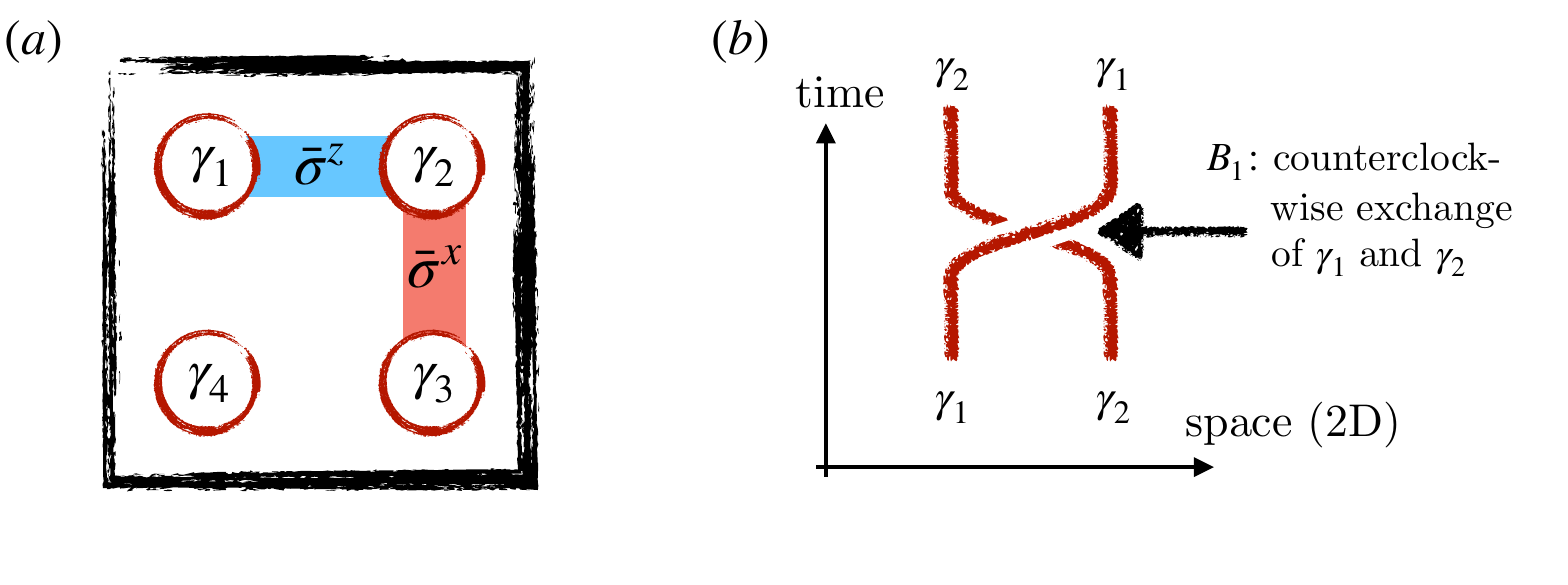}
  \caption{%
  $(a)$ Sketch of the parity Majorana qubit: Two Majorana zero modes together
  form a single Dirac fermionic mode whose Hilbert space is two-dimensional as
  the mode can either be empty or filled. Both states have the same energy.
  Four Majorana zero modes thus form a four-dimensional Hilbert space of
  which, due to the conservation of the total fermion parity, only a
  two-dimensional subspace can be accessed. This degenerate, two-dimensional
  subspace is the Majorana qubit. Gates on the qubit can be  performed either
  by braiding or by coupling two Majorana zero modes. As indicated in the
  figure, coupling $\gamma_1$ to $\gamma_2$ implements a $\bar
  \sigma^z$-operation, whereas coupling $\gamma_2$ to $\gamma_3$ leads to a
  $\bar \sigma^x$-operation.  Given the fact that the Majorana zero modes are
  sufficiently far apart from each other and that the environment only acts
  locally on the system, these operations are not performed `accidentally' by
  the environment and the Majorana qubit is protected from both sign flip and
  bit flip errors. As this protection originates from the conservation of the
  total fermion parity, the qubit is called parity-protected.  $(b)$
  Elementary operation of the braid group. The geometric representation of the
  braid group is in space-time; the horizontal axis is the spatial axis
  whereas the vertical one is temporal.  The counterclockwise exchange of
  Majorana zero modes $\gamma_1$ and $\gamma_2$ in space-time forms the braid
  $B_1$.
  }\label{fig:parity_qubit}
\end{figure}
We have seen in the last section that due to the parity-conservation, we need
to have two fermionic modes to encode a single qubit. For concreteness, we
will work in the even parity superselection sector and have the single logical
qubit encoded as $|\bar 0 \rangle=|00\rangle$ and $|\bar 1\rangle=|11\rangle$.
Thinking about a possible implementation in terms of Majorana  modes, we
encode each fermionic mode in a pair of Majorana zero modes which are
localized states sufficiently far separated from each other, see
Fig.~\ref{fig:parity_qubit}. We denote the Majorana zero modes on the top as
$\gamma_{1}$ and $\gamma_{2}$ , and the ones on the bottom as $\gamma_3$ and
$\gamma_4$. The Majorana modes are at zero energy thus the two states $|\bar 0
\rangle$ and $|\bar 1 \rangle$ are degenerate in energy. The parities of the
number of electrons on the superconducting segments are given by
$\mathcal{P}_{1} = -i \gamma_{1}\gamma_{2}$ and $\mathcal{P}_2= -i \gamma_3
\gamma_4$. Due to the parity constraint, we have $\mathcal{P}_1 =
\mathcal{P}_2$ and the action of both operators on the logical qubit emulates
the (logical) Pauli-operator
\begin{equation}\label{eq:sigma_z}
  \bar\sigma^z = - i \gamma_{1}\gamma_{2} =  - i \gamma_{3}\gamma_{4}\, .
\end{equation}

In order to have a complete qubit, we are left with the task of finding
a logical $\bar\sigma^x$, an operator which anticommutes with $\bar\sigma^z$.
It is easy to see that
\begin{equation}\label{eq:sigma_x} 
  \bar \sigma^x =-i \gamma_{2} \gamma_{3}
  =- i \gamma_{1} \gamma_{4}
\end{equation}
anticommutes with $\bar\sigma^z$ due to the fact that the single Majorana
fermions shared by both operators anticommute with one another. In the
situation where all the Majorana modes are sufficiently far separated from
each other, any gate on the logical qubit is a nonlocal operator. Due to this
nonlocality, it is highly unlikely that uncontrolled, random fluctuations in
the environment will execute a gate and thus cause an error on the logical
qubit. This protection of the Majorana qubit is called symmetry-protected
topological order \cite{gu:09,pollmann:12} or simply parity-protection
\cite{hassler:11}. The decisive difference to full topological order, as it is
for example present in Kitaev's toric code \cite{kitaev:06}, is the fact that
logical Pauli operators are only required to be nonlocal as long as the parity
symmetry is conserved. A tunneling of single electrons from a reservoir onto
the superconducting island is a local process that violates parity
conservation, immediately bringing the Majorana qubit out of its computational
subspace.

The requirement for operating the Majorana qubit successfully in a protected
manner is that the environment does not provide single unpaired electrons.
This sounds very stringent at first sight. However, the physical
implementation of the system only involves superconductors where most of the
electrons are paired up into Cooper pairs and where, at temperature $T$, only
an exponentially small fraction proportional to the Boltzmann factor
$e^{-\Delta/k_B T}$ remains unpaired.  The storage time of quantum information
in a Majorana qubit thus is expected to increase exponentially when lowering
the electron temperature.

\subsection{Fusion and splitting}

\begin{figure}
  \centering
  \includegraphics[width=0.8\textwidth]{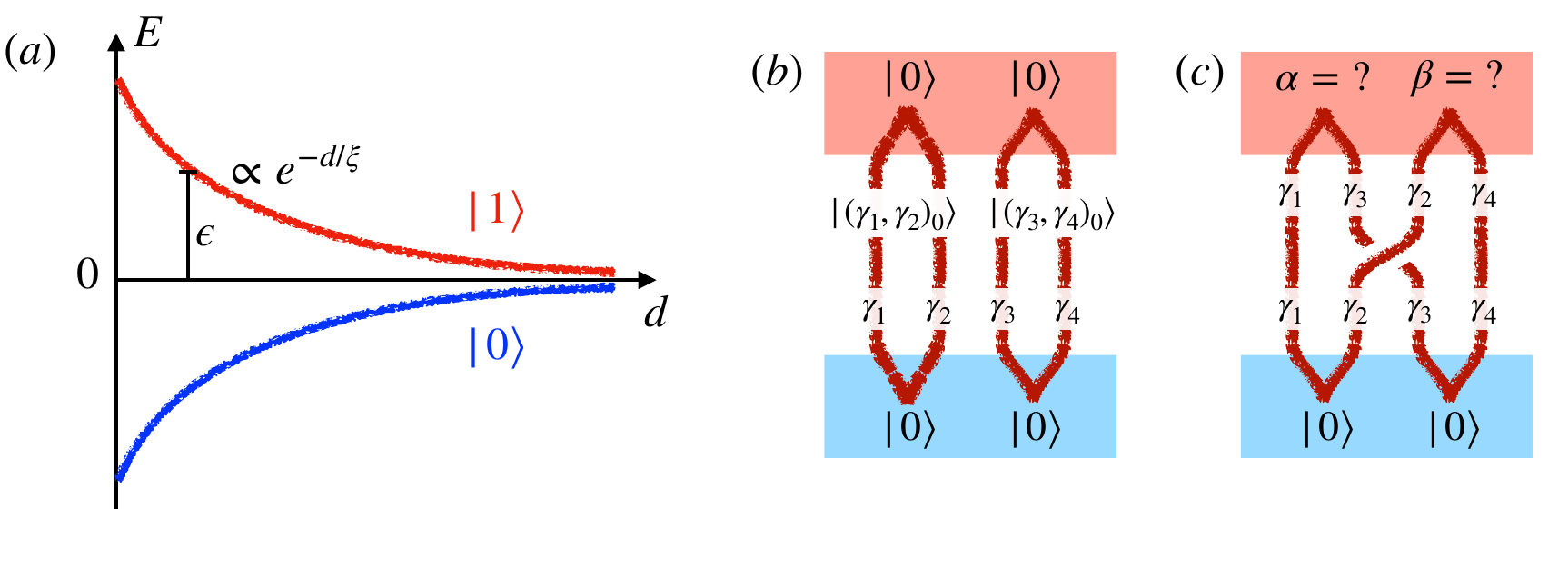}
  \caption{%
    $(a)$ Bringing two Majorana zero modes in close proximity (called fusion)
    such that their distance $d$ is of the decay length $\xi$, the energy
    difference $\propto \exp(-d/\xi)$ between the state $|0\rangle$ and
    $|1\rangle$ breaks the parity protection. By measuring whether or not the
    system is in the ground state, a projective measurement in the
    computational basis states $|0\rangle$ and $|1\rangle$ is performed.
    Similarly, we can split the Majorana zero modes that are initialized in
    the ground state $|0\rangle$ and produce a parity protected state. $(b)$
    The fusion outcome (red box) of Majorana zero modes that have been
    initialized to the vacuum (by splitting, blue box) is well-defined. $(c)$
    If the Majorana zero modes have been interchanged (braided), the fusion
    outcomes $\alpha,\beta$ fluctuate.
  }\label{fig:fusion}
\end{figure}

We have seen before that if you bring the Majorana zero modes $\gamma_1$ and
$\gamma_2$ close together, there is a finite energy splitting $\epsilon \simeq
\bar \mu e^{-d/\xi}$ due to the overlap of the modes, see
Fig.~\ref{fig:fusion}$(a)$. This splitting breaks the parity protection and as
a result the two states $|0\rangle$ and $|1\rangle$ are no longer degenerate
in energy. It is then possible to detect in which state the system is
\cite{nayak:08,stern:08,hassler:11}.  The process of bringing two Majorana
modes together is called fusion. Graphically, we denote the fusion experiment
as the red box in Fig.~\ref{fig:fusion}$(b)$.  The fusion experiment
implements the measurement in the computational basis and is the last step in
a quantum computation.\index{fusion}

The reverse process is called splitting and is depicted as the blue box in
Fig.~\ref{fig:fusion}$(b)$. If one starts with the vacuum state $|0\rangle$
when the Majorana zero modes are overlapping, the Majorana states can be
separated from each other such that the initial state $|0\rangle$ becomes
degenerate with $|1\rangle$ and parity protection is achieved. It is clear
that if one fuses the zero modes again immediately after the splitting, the
outcome of the fusion experiment is the state $|0\rangle$ with certainty, see
Fig.~\ref{fig:fusion}$(b)$. On the other hand, if the pairing is changed such
that at the fusion different Majorana zero modes are paired up as in the
splitting, the result is unclear, see Fig.~\ref{fig:fusion}$(c)$.

As we are now thinking about different ways of pairing the Majorana zero modes
to produce a Dirac fermion that can be either empty or occupied, it is useful
to introduce the notation $|(\gamma_1, \gamma_2)_j\rangle$. It denotes whether
the state formed by $\gamma_1$ and  $\gamma_2$ is occupied $(j=1)$ or empty
$(j=0)$. With this new notation, we can write for the logical states
\begin{equation}
  |\bar 0 \rangle = \bigl| (\gamma_1, \gamma_2)_0\; (\gamma_3, \gamma_4)_0
  \bigr\rangle ,
  \qquad  |\bar 1 \rangle = 
  \bigl| (\gamma_1, \gamma_2)_1 \;(\gamma_3, \gamma_4)_1
  \bigr\rangle .
\end{equation}
The fusion experiment in Fig.~\ref{fig:fusion}$(c)$ produces the outcome
$\alpha, \beta\in \{0,1\}$ with probability
\begin{equation}\label{eq:pab}
  P(\alpha,\beta) = \left|\bigl\langle (\gamma_1,\gamma_3)_\alpha \;(\gamma_2,
  \gamma_4)_\beta \big| (\gamma_1, \gamma_2)_0\; (\gamma_3, \gamma_4)_0 
  \bigr\rangle
  \right|^2\,.
\end{equation}
Due to the conservation of total parity $\mathcal{P}$, the only possible
outcomes are $\alpha = \beta =0$ and $\alpha = \beta =1$. We will calculate
the probabilities in the next section. However, we must first introduce the
part of Fig.~\ref{fig:fusion}$(c)$  in between the splitting and the fusion of
the Majorana zero modes. The exchange of the zero modes is called
braiding.\index{braiding}

\section{Braiding}

\begin{figure}[tb]
  \centering
  \includegraphics[width=0.9\linewidth]{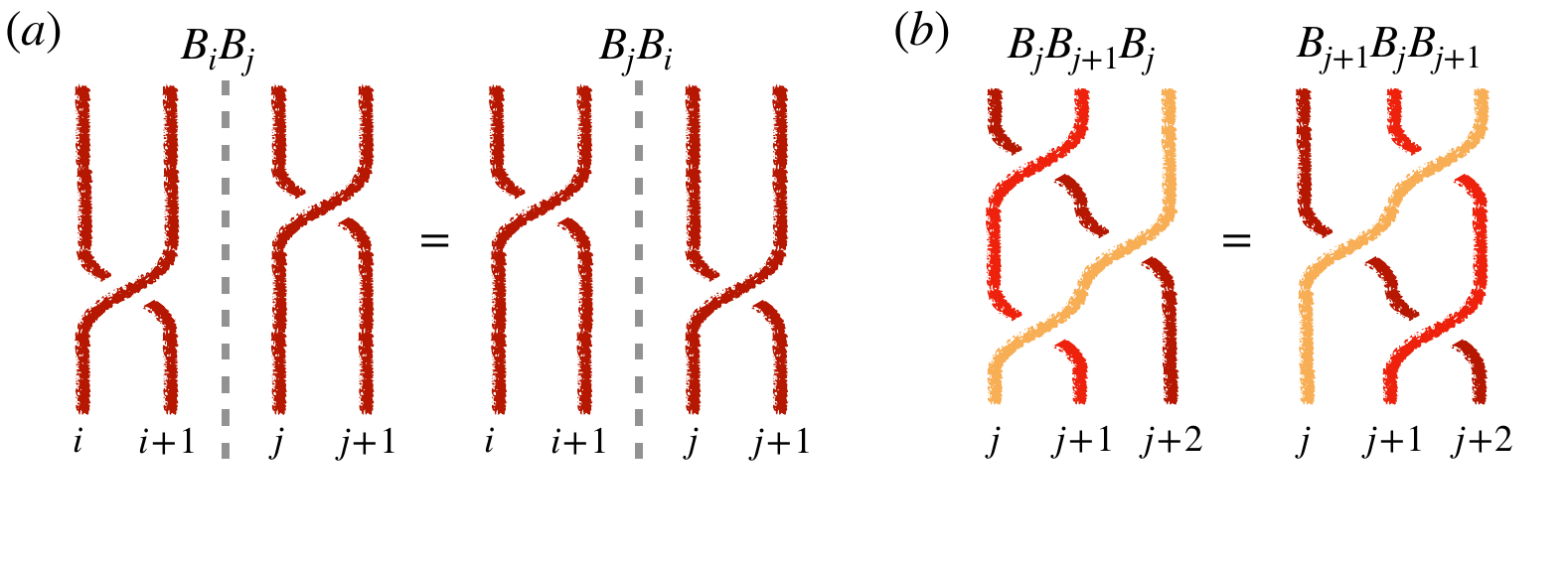}
  \caption{%
  $(a)$ Braiding operators commute if their index $i$ and $j$ are more than 2
  apart. The reason is that $B_i$ and $B_j$ do not have any strand in common
  so either one can be executed first.  $(b)$ The Yang-Baxter equation $B_j
  B_{j+1} B_j = B_{j+1} B_j B_{j+1}$ provides a nontrivial relation between
  the generators $B_j$ and $B_{j+1}$.  That the two braids are topologically
  equivalent can be seen as follows: in both braids the dark strand can be
  considered to lie in the very back and connects the initial position $j+2$
  to the final position $j$. Similarly, the bright strand lies in front and
  connects $j$ to $j+2$.  The middle strand starts at $j+1$ and ends at the
  same place. The braids are equivalent as they can be deformed into each
  other by sliding the middle strand $j+1$ in between the two other strands
  from the left to the right.
  }\label{fig:yb}
\end{figure}

Exchange statistics is introduced in the basic physics courses as the action
of permutations of the symmetric group  $\mathcal{S}_N$ with $N!$ elements on
the wavefunction of $N$ identical particles.\index{identical particles}
There are two possibilities: either the wavefunction remains invariant
(`bosons') or the wavefunction acquires a minus sign (`fermions') under the
exchange of two particles. On a mathematical level, the origin of this
distinction lies in the fact that the Hamiltonian of identical particles
commutes with an arbitrary element of the symmetric group $\mathcal{S}_N$.
Thus, it is possible to classify the eigenstates of the Hamiltonian in terms
of irreducible representations of the permutation group. There are only two
one-dimensional representations: the trivial representation (corresponding to
bosons) and the sign representation (corresponding to fermions).  Any
representation whose dimension is larger than one leads to a degeneracy, which
is called exchange degeneracy as it originates simply from the fact that
particles are indistinguishable.%
\footnote{The statistics of identical particles that transforms according to
higher dimension representations of the permutation groups is called
parastatistics.  However, even if particles with parastatistics were to exist,
they would offer nothing new as a set of Klein transformations could be used
to map particles with parastatistics onto bosons or fermions with a set of
internal quantum numbers (like spin, \dots).  Later, we will see that such a
mapping is not possible in $2+1$ dimensions and that higher dimensional
representations of the braid group is truly different from the
one-dimensional representations.}
In fact, the spin-statistics
theorem can be proven in the context of relativistic field theory in $3+1$
dimensions, which states that particles with integer spin are bosons whereas
particles with half-integer spin are fermions.\index{braid group}

It has been pointed out  that the physical process of exchanging two particles
is important. In fact, the exchange of two identical  particles  has to be
viewed as a (slow) process that occurs in space-time \cite{wilzcek:82}.  In
$2+1$ dimensions, the relevant group is the braid group $\mathcal{B}_N$ of $N$
strands as trajectories in space-time for exchanging two particles clock- or
counterclockwise are topologically distinct.  The braid group $\mathcal{B}_N$
consists of $N$ strands. The generators of the group $B_j$ denote the braiding
strand $j$ and $j+1$ in the counterclockwise direction (in
Fig.~\ref{fig:parity_qubit}$(b)$, $B_1$ braids strand 1 and 2
counterclockwise). The inverse operation braids the strands clockwise. Two
elements of the group are equivalent if the corresponding braids can be
smoothly deformed into each other by keeping their ends fixed.  Note that
different from the symmetric group $B_j \neq B_j^{-1}$.  The braid group of
$N$ strands is generated by $B_1, \dots, B_{N-1}$ where the generators satisfy
the following two relations
(Artin\cite{artin:47}),\index{braiding}\index{Yang-Baxter equation}
\begin{equation}\label{eq:braid_group}
  B_i B_j = B_j B_i, \quad |i-j|\geq 2
  \qquad\text{and}\qquad
  B_j B_{j+1} B_j = B_{j+1} B_j B_{j+1};
\end{equation}
the latter is also called Yang-Baxter equation, see Fig.~\ref{fig:yb}.
Unlike the symmetric group $\mathcal{S}_N$ the group order is infinity
which makes the classification of all irreducible representations difficult.

In $3+1$ dimensions, clockwise and counterclockwise depends on the observer
(coordinate system) and thus the exchanges $B_j$ and $B_j^{-1}$ are
topologically equivalent. As a result, we obtain $B_j^2 = B_{j}^{-1} B_j =1$.
Under this additional constraint, the braid group $\mathcal{B}_N$ reduces to
the symmetric group $\mathcal{S}_N$. The dependence of the exchange statistic
on the dimension of space is as follows:\index{exchange statistics}
\begin{itemize}
  \item 1D: \quad no exchange of particles is possible
  \item 2D: \quad the exchange is described by the braid group $\mathcal{B}_N$
  \item 3D: \quad we have $B_j^2 = 1$ and the symmetric group $\mathcal{S}_N$
    characterizes the exchange (only bosons and fermions are possible)
\end{itemize}
We see that 2D is special. Thus, we will study some examples of nontrivial
exchange statistics in $2+1$ dimensions in the following.

\subsection{Abelian anyons}

\index{anyon!Abelian}

The one-dimensional (unitary) representations of the braid group are simple to
construct. In general,  the action of $B_j$ onto a wavefunction is given by a
phase factor $\rho(B_j) = e^{i\theta_j}$ with $\theta_j \in [0,2\pi)$. The
Yang-Baxter equation demands
\begin{equation*}
  \rho(B_j) \rho(B_{j+1}) \rho(B_j) = \rho(B_{j+1}) \rho(B_j) \rho(B_{j+1})
  \;\Rightarrow \; e^{2i \theta_j + i \theta_{j+1}} = e^{i \theta_j +2 i
  \theta_{j+1}} \; \Rightarrow \; \theta_j = \theta_{j+1} 
\end{equation*}
for any 1D representation. As a result, all the angles are equal and the
representation
\begin{equation}
  \rho_\theta(B_j) =e^{i\theta}
\end{equation}
is characterized by a single angle $\theta$.

Note that for $\theta=0$, we get the customary result for bosons that
interchanging two particles does nothing to the wavefunction whereas for
$\theta=\pi$ interchanging introduces a minus sign which is the result for
fermions. In $2+1$ dimensions, all angles in between 0 and $\pi$ are allowed
and particles with $\theta \neq 0$ or $\pi$ are called (Abelian) anyons. As an
example, we note that quasiparticles in the fractional quantum Hall effect at
filling fraction $\nu = \tfrac1n$ with $n$ an odd integer are anyons with
$\theta = \nu \pi$. Even though the Abelian anyons are interesting from a
physical point of view, they do not offer any resources that can be used for
quantum computing as braiding only introduces simple phase factors.

Particles whose wavefunctions transform according to higher dimensional
irreducible representations of the braid group are called non-Abelian anyons.
A necessary ingredient is a ground state degeneracy (which grows exponentially
with the number of particles). The effect of $B_j$, the counterclockwise
exchange of two particles $j$ and $j+1$, is then represented by a unitary matrix
$\rho(B_j)$ on the ground state manifold. As different unitary matrices do not
commute, the representation is non-Abelian which is the reason for their name.

The usefulness of non-Abelian anyons for topological quantum computation
relies on the fact that the degeneracy of the ground state manifold is
protected, and the gates implemented by the exchange of particles are exact (up
to an unimportant global phase) \cite{nayak:08}. A specific species of
non-Abelian anyons is called universal for quantum computation, if, for any
given gate, a braid can be found which approximates the gate with arbitrary
accuracy.

\subsection{Ising anyons}

\index{anyon!non-Abelian}
We have seen that Majorana zero modes in topological superconductors lead to a
degenerate ground state of size $2^{N-1}$ that grows exponentially with the
number of zero modes. It is thus a natural question to ask if the process of
braiding the zero modes lead to a nontrivial operation $\rho_\gamma(B_j)$ on
the ground state manifold. It can be shown, see the Appendix, that
braiding of Majorana zero modes in topological superconductors is described by
the unitary representation\cite{read:00,ivanov:01,alicea:11,clarke:11,heck:12}
\begin{equation}
  \rho_{\gamma}(B_j) = \exp\left(- \frac\pi4 \gamma_j \gamma_{j+1}\right) =
  \frac1{\sqrt2} (1 - \gamma_j \gamma_{j+1})
\end{equation}
of the braid group. The representation $\rho_\gamma$ is also called Ising
anyons.  The clockwise exchange of the strands $j$ and $j+1$ is implemented by
$\rho_\gamma (B_j^{-1}) = \rho_\gamma (B_j)^\dag= \rho^\dag_\gamma (B_j) =
\exp\left(\frac\pi4 \gamma_j \gamma_{j+1}\right) $.

Under the action of $\rho_\gamma(B_j)$, the Majorana zero modes are mapped
onto each other with
\begin{equation}
  \gamma_j' = \rho^\dag_\gamma(B_j) \gamma_j \rho^\pdag_\gamma(B_j) 
  =- \gamma_{j+1}, \qquad
  \gamma_{j+1}' = \rho^\dag _\gamma(B_j)
  \gamma_{j+1} \rho^\pdag_\gamma(B_j) = \gamma_j
\end{equation}
while the other modes remain unaffected.  That this is the correct expression
is reinforced by the fact that the (local) parity
\begin{equation}
  \mathcal{P}'_j=
  \rho^\dag_\gamma(B_j) \mathcal{P}_j \rho^\pdag_\gamma(B_j) = - i \gamma'_j
  \gamma'_{j+1} = -i \gamma_j \gamma_{j+1} = \mathcal{P}_j
\end{equation}
is conserved\cite{halperin:12}. 

To check that $\rho_\gamma$ is a representation of the braid group, we have to
check $\rho_\gamma(B_i) \rho_\gamma(B_j) = \rho_\gamma(B_j) \rho_\gamma(B_i)$,
$|i-j|\geq 2$, and the Yang-Baxter equation, see Eq.~\eqref{eq:braid_group}.
The first relation follows easily as $\rho_\gamma(B_i)$ and $\rho(B_j)$
commute due to the fact that they act on different Majorana zero modes and
consist of an even number of Majorana operators. The Yang-Baxter equation can
be directly evaluated
\begin{align}
  \rho_\gamma(B_j) \rho_\gamma(B_{j+1})  \rho_\gamma(B_{j}) &= \frac1{2^{3/2}}
  ( 1- \gamma_{j} \gamma_{j+1})( 1- \gamma_{j+1} \gamma_{j+2})  ( 1- \gamma_j
  \gamma_{j+1})  \nonumber \\
  &=- \frac{1}{\sqrt{2}} \left(\gamma_j \gamma_{j+1} +
  \gamma_{j+1} \gamma_{j+2}\right)\,,
\end{align}
and similarly
\begin{equation}
  \rho_\gamma(B_{j+1}) \rho_\gamma(B_{j})  \rho_\gamma(B_{j+1})=-
  \frac{1}{\sqrt{2}} \left(\gamma_j \gamma_{j+1} +
  \gamma_{j+1} \gamma_{j+2}\right)\,;
\end{equation}
as a result, the unitary gate performed by braiding Majorana zero modes only
depends on the braid and not on the specific paths taken. In particular, both
methods depicted in Fig.~\ref{fig:yb}$(b)$ of moving the Majorana zero modes
around each other produces the same operation on the ground state manifold.

Thus, braiding can be used to perform topologically protected gates onto the
encoded Majorana qubit states $|\bar 0 \rangle$ and $|\bar 1 \rangle$. Indeed,
for $N=4$, the interchange of Majorana zero modes implements the operations
\begin{equation}
  \rho_\gamma(B_1) = \rho_\gamma(B_3) 
  = \exp\left(-i\frac{\pi}4 \bar\sigma^z\right),
    \qquad \rho_\gamma(B_2) =\exp\left(-i\frac{\pi}4 \bar\sigma^x\right)= 
    \frac{1}{\sqrt{2}} (1 -i \bar \sigma^x)
\end{equation}
that correspond to rotations by $90^\circ$ on the Bloch sphere. 

We  now have all the ingredients to calculate the probability
$P(\alpha,\beta)$ of Eq.~\eqref{eq:pab}, see Fig.~\ref{fig:fusion}$(c)$.
Performing the braid $B_2$ acts as $\rho_\gamma(B_2) = 2^{-1/2} (1 -i \bar
\sigma^x)$ on the ground state manifold. In particular, the initial state
$|\bar 0 \rangle = \bigl|(\gamma_1,\gamma_2)_0\;(\gamma_3,\gamma_4)_0
\bigr\rangle$ gets transformed into
\begin{equation}
  \rho_\gamma (B_2)|\bar 0 \rangle = \frac1{\sqrt2} (|\bar 0 \rangle  
  - i |\bar
  1 \rangle)  = \frac1{\sqrt2} (| 00 \rangle  - i |11 \rangle) \,.
\end{equation}
The probabilities $P(\alpha,\beta)$ for the outcomes $\alpha = \beta \in
\{0,1\}$ after  the braid are given by
\begin{equation}
  P(0,0) = \bigl| \langle 0 0 |  \rho_\gamma (B_2)| \bar 0 \rangle \bigr|^2 =
  \frac12, \qquad  P(1,1) = 
  \bigl| \langle 1 1 |  \rho_\gamma (B_2)| \bar 0 \rangle \bigr|^2 =
  \frac12.
\end{equation}
As the result $P(\alpha,\alpha)=50\,\%$ for both $\alpha =0$ and $\alpha=1$
originates from the representation of the braid group of the Majorana zero
modes, the fusion experiment in Fig.~\ref{fig:fusion}$(c)$ has been proposed
recently as a first test of the non-Abelian nature of Majorana zero modes
\cite{aasen:16}.

Even though these operations are protected, braiding of Ising anyons is not
enough to perform arbitrary unitary operations on the ground state manifold.
The single qubit rotations are not complete as only rotations by (multiples
of) $90^\circ$ around the coordinate axes $x,y,z$ can be implemented.  In
particular, a rotation by $45^\circ$ is missing (called $\frac\pi8$-phase or
T-gate) and an entangling gate needs to be added\cite{bravyi:05,nayak:08}. In
concrete realizations, entanglement can be obtained by performing a joint
parity measurement of two qubits \cite{hassler:11,hassler:13}. In fact, it is
enough if the T-gate is implemented with a fidelity of 90\% as a distillation
protocol using the exact Clifford gates in a process called \emph{Magic state
distillation} can be employed to purify the state
\cite{bravyi:05}.\index{magic state distillation}

\subsection{Fibonacci anyons}

The Ising anyons provided by the Majorana modes have the nice property that
they allow for noise-insensitive operations on a parity-protected qubit  by
braiding. Still, the group of operations that can be obtained in this way is
not enough for universal quantum computation. However, luckily, there exist
other anyons, in particular Fibonacci anyons, that allow for a universal set
of operations by braiding \cite{freedman:02,nayak:08}.  That Majorana zero
modes do not offer universality is connected to the fact that the Hilbert
space can be locally assigned to Majorana modes: any pair of Majorana zero
modes can either be filled or empty. For Fibonacci anyons the relation between
the anyons and the Hilbert space of the ground state thus has to be more
complicated to overcome this issue.

We denote Fibonacci anyons with the letter $\tau$. A pair of Fibonacci anyons
has two possible fusion outcomes: it can either fuse to the vacuum 0 or to a
Fibonacci anyon $\tau$. In particular, we have that $|(\tau_1,
\tau_2)_0\rangle$, $|(\tau_1, \tau_2)_\tau\rangle$ forms a basis of the
two-dimensional ground state manifold of two Fibonacci anyons. However, akin
to the Ising anyons, there is a superselection rule forbidding us to use these
two states as a qubit. The `Fibonacci qubit' thus has to be formed from a
subspace of three Fibonacci anyons $\tau_1, \tau_2, \tau_3$ with the possible
fusion outcomes\footnote{Note that the fusion of the vacuum $0$ with $\tau$
can only give $\tau$.}\index{qubit!Fibonacci}
\begin{equation}
  |\bar0\rangle = |((\tau_1, \tau_2)_0, \tau_3)_\tau\rangle,
  \quad |\text{nc}\rangle=
 |((\tau_1,
  \tau_2)_\tau,   \tau_3)_0\rangle, \quad
  |\bar 1\rangle = |((\tau_1, \tau_2)_\tau,
  \tau_3)_\tau\rangle\,.
\end{equation}
Note that the first and the last state are in the same superselection sector
(as they fuse to $\tau$) and thus can be used as a genuine Fibonacci qubit.
The state of the qubit is then determined by the fusion outcome of the first
two Fibonacci anyons while the fusion with the last Fibonacci anyon is
constrained to be $\tau$. The third state with total fusion outcome $0$ is
then a noncomputational (nc) state.

For the Hilbert space of the degenerate ground state of the $N$-Fibonacci
anyons $\tau_1, \dots, \tau_N$, we introduce the following notation
\begin{equation}\label{eq:fib_basis}
  |0,\tau,f_2,\dots,f_N\rangle= |({\scriptstyle \dots}
  ((\tau_1, \tau_2)_{f_2}, \tau_3)_{f_3}, \dots )_{f_N} \rangle, \qquad f_j
  \in \{0,\tau\},
\end{equation}
where we have added the fusion outcomes $0$ and $\tau$ of the zeroth and first
Fibonacci anyons for future convenience.

A particular state is then labeled by writing the fusion outcome $f_j$ in
between strand $j$ and $j+1$. In order to determine the dimension of the
ground state, we have to find the number of states of the form in
Eq.~\eqref{eq:fib_basis}. The fusion outcomes $f_0=0,f_1, \dots,f_N$ are only
constrained by the fact that the fusion of the vacuum $f_j=0$ (at step $j$)
with a Fibonacci anyon necessarily gives a Fibonacci anyon as the fusion
outcome and thus $f_{j+1} =\tau$. Thus, we need to count the number of states
$|0,\tau,f_1,\dots,f_N\rangle$ without two 0s in a row.  Denoting by $Z_j$
($O_j$) the number of states with $f_j=0$ ($f_j=1$), we have to solve the
recurrence relations
\begin{align}
  Z_{j+1} = O_j, & \quad O_{j+1} = O_j + Z_j,  \qquad O_{0}=0, \quad Z_0 =1\,,
\end{align}
with the result $Z_{j+1} =  O_j = F_j$, $F_j$ being the $j$-th Fibonacci
number.  Note that this relation is the reason why the anyons are called
Fibonacci anyons.

\begin{table}
\parbox[c]{.2\textwidth}{%
  \centering
\includegraphics[width=.7\linewidth]{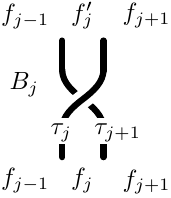}\vspace{10mm}}%
  \parbox[c]{.8\textwidth}{
  \centering
    \[
   \renewcommand\arraystretch{1.3}
  \begin{array}{lcl}
    \hline
    |i\rangle = |f_{j-1}, f_j, f_{j+1}\rangle &\qquad & 
    \rho_\tau(B_j) |i\rangle = \sum_{f'_j} c_{f'_j}
    |f^{\vphantom '}_{j-1}, f'_j , f^{\vphantom '}_{j+1}\rangle
    \\ \hline
    |0,\tau,\tau \rangle & & \omega^{-1} | 0,\tau,\tau \rangle \\
    |\tau,\tau,0\rangle & &  \omega^{-1} | \tau,\tau, 0 \rangle \\
    |0,\tau,0\rangle & &  \omega^{-2} | 0,\tau, 0 \rangle \\
    |\tau,0,\tau \rangle  & & \phi^{-1} \omega^2 |\tau,0,\tau\rangle 
    + \phi^{-1/2} \omega
    | \tau,\tau,\tau \rangle \\
    | \tau,\tau,\tau \rangle && \phi^{-1/2} \omega
    | \tau,0,\tau \rangle   - \phi^{-1} | \tau,\tau,\tau \rangle
    \\\hline\hline
  \end{array}
\]}
\caption{Result of the elementary braid $B_j$ of  Fibonacci anyons on the
basis formed by the fusion outcomes. Note that the effect of braiding the
strands $j$ and $j+1$ is local in the sense that it only changes the fusion
outcome $f_j$.  The parameters are the phase $\omega = - q =- e^{2 \pi i/5}$
and the golden ratio $\phi = (1+\sqrt 5)/2 = q + \bar q +1$.}\label{tbl:braid}
\end{table}

The unitary representation  $\rho_\tau(B_j)$ of braiding the Fibonacci anyons
$\tau_j$ and $\tau_{j+1}$ can be expressed as a local relation of the fusion
outcomes $f_{j-1}, f_j, f_{j+1}$, see Table~\ref{tbl:braid} and
\cite{kauffman:07,shor:08}.\footnote{Note that the references denote our
$0,\tau$ by $*,p$.} It can be directly checked that $\rho_\tau$ is unitary. To
check that it is, in fact, a representation, we have to prove
Eq.~\eqref{eq:braid_group}.  As before, the first relation in
\eqref{eq:braid_group} follows from the fact that $\rho_\tau(B_j)$ only
depends on $f_{j-1}, f_j, f_{j+1}$ and only changes the value of $f_j$. As a
result, $\rho(B_i)$ trivially commutes with $\rho(B_j)$ for $|i-j|\geq 2$.

In the last step,  we have to test whether  $\rho_\tau$ satisfies the
Yang-Baxter equation. As for the Ising anyons, we do this directly by
calculating $U V U$ and $VUV$ with $U= \rho_\tau(B_j)$ and $V=
\rho_\tau(B_{j+1})$ separately and subsequently verify that $UVU=VUV$. The
Yang-Baxter equation acts on the three strands  $j$, $j+1$, $j+2$.  In the
first step, we determine the representation of $\rho_\tau(B_j)$ and
$\rho(B_{j+1})$ in the Hilbert space $|f_{j-1}, f_j, f_{j+1}, f_{j+2}\rangle$.
Note that we suppress the labels of the other fusion outcomes as they remain
unchanged.  Moreover, we see from Table~\ref{tbl:braid} that $U$ and $V$
cannot change the value of  $f_{j-1}$ and $f_{j+2}$ (superselection rule)  and
thus we can verify the relation for each value of  $a=f_{j-1}$ and $b=f_{j+2}$
separately.

With the rules of Table~\ref{tbl:braid}, we find [$U_{ab}$ is the matrix
$\rho(B_j)$ in the subspace with fixed $a$ and $b$]
\begin{align}\label{eq:uandv}
  &U_{00}=V_{00}=(\omega^{-1}); & &\!U_{0\tau}=
  \begin{pmatrix} \omega^{-2} & 0 \\
  0  & \omega^{-1}\end{pmatrix}, V_{0\tau} =   
  \begin{pmatrix} \phi^{-1}\omega^2 & \phi^{-1/2}\omega \\
  \phi^{-1/2} \omega & -\phi^{-1}\end{pmatrix}; \\
  &\!U_{\tau0} =V_{0\tau}, V_{\tau0}= U_{0\tau}; &
  & U_{\tau\tau} = \begin{pmatrix}
    \phi^{-1}\omega^2 &0  &\phi^{-1/2}\omega \\
    0 &\omega^{-1} &0\\
  \phi^{-1/2} \omega & 0& -\phi^{-1}\end{pmatrix},
  V_{\tau\tau} = \begin{pmatrix}
    \omega^{-1} &  0& 0 \\
 0 & \phi^{-1}\omega^2 & \phi^{-1/2}\omega \\
  0& \phi^{-1/2} \omega & -\phi^{-1}\end{pmatrix} .
    \nonumber
\end{align}
Here, we have ordered the basis states as follows:
$\{|0,\tau,\tau,0\rangle\}_{00}$, $\{|0,\tau,0,\tau\rangle,
|0,\tau,\tau,\tau\rangle \}_{0\tau}$, $\{|\tau,0,\tau,0\rangle,
|\tau,\tau,\tau,0\rangle\}_{\tau 0}$, and $\{|\tau,0,\tau,\tau\rangle,
|\tau,\tau,0,\tau\rangle, |\tau,\tau,\tau,\tau\rangle\}_{\tau\tau}$. It is now
an easy exercise in matrix multiplication to verify $U V U = V U V$ and thus
to show that $\rho_\tau$ is a unitary representation of the braid group.

We obtain more insights into the operations performed by braiding by choosing
$j=1$, $f_0=0$ and looking at the states $|((\tau_1,\tau_2)_a,
\tau_3)_b\rangle$ of the three Fibonacci anyons in Eq.~\eqref{eq:fib_basis}.
Performing $B_1$ by braiding the first two anyons (corresponding to the
matrices $U_{0b}$), the states simply acquire the phases $\omega^{-2} =
e^{-4\pi i/5}$ (if $a=0$) and $\omega^{-1} =e^{3\pi
  i/5}$ (if $a=\tau$) depending only on their fusion outcome $a$, irrespective
of the value of $b$.\footnote{The corresponding phases are known as elements
of the R-matrix.}

On the other hand, braiding the anyons $\tau_2$ and $\tau_3$ with $B_2$
(corresponding to $V_{0b}$) is not diagonal in this basis as the two anyons do
not have a well-defined fusion outcome; in the alternative basis $|(\tau_1,
(\tau_2,\tau_3)_{a'})_b\rangle$ the operation $B_2$ would be the simple phase
factors $\omega^{-1},\omega^{-2}$ depending on $a'$ as before. It is easy to
check that $V_{0\tau} = F^{-1} U_{0\tau} F$ with the transformation matrix
\begin{equation}
  F= F^{-1}= \begin{pmatrix} \phi^{-1} & \phi^{-1/2} \\
  \phi^{-1/2} &-\phi^{-1}\end{pmatrix},
\end{equation}
which corresponds to the basis transformation
\begin{align}
  | ( \tau_1, (\tau_2, \tau_3)_\tau )_0 \rangle&=  | ( (\tau_1,\tau_2)_\tau,
  \tau_3)_0 \rangle, \\ 
  | ( \tau_1, (\tau_2, \tau_3)_0 )_\tau \rangle&=
  \phi^{-1}  | ( (\tau_1,\tau_2)_0, 
  \tau_3)_\tau \rangle  +  \phi^{-1/2}  | ( (\tau_1,\tau_2)_\tau,
  \tau_3)_\tau \rangle, \nonumber \\
| ( \tau_1, (\tau_2, \tau_3)_\tau )_\tau \rangle&=
  \phi^{-1/2}  | ( (\tau_1,\tau_2)_0, 
  \tau_3)_\tau \rangle  -  \phi^{-1}  | ( (\tau_1,\tau_2)_\tau,
  \tau_3)_\tau \rangle, 
\end{align}
also known as the $F$-move.\index{$F$-move}

A single Fibonacci qubit is realized by three Fibonacci anyons $\tau_1,
\tau_2, \tau_3$ in the state $|0,\tau, f, \tau\rangle$. The states with
$f\in\{0,\tau\}$ form the logical qubit $|\bar 0\rangle, | \bar 1\rangle$.
From the calculation in Eq.~\eqref{eq:uandv} we know that the gate $U_{0\tau}$
with the phases $e^{-\pi i/10 \pm 7\pi i/10}$ is performed while braiding the
first two anyons. Up to the irrelevant (Abelian) factor,  $e^{-\pi i/10}$, the
braid corresponds to a rotation by $\frac{7}5\pi \equiv252^\circ$ around the
$z$-axis of the Bloch sphere.  Similarly, braiding anyons $\tau_2$ and
$\tau_3$ yields the gate $V_{0\tau}$ in a protected fashion. This gate is
another rotation by $252^\circ$ around an axis that corresponds to the basis
transformation given by $F$. In terms of the Bloch sphere, the rotation
$V_{0\tau}$ is a rotation around the axis $\mathbf v =\bigl(2 \phi^{-3/2} ,0,
\phi^{-2}- \phi^{-1}\bigr)^T$.  Braiding the anyons $\tau_1, \tau_2, \tau_3$,
arbitrary products of $U_{0\tau}$ and $V_{0\tau}$ can be implemented and all
single qubit gates can be performed in a protected manner.

\begin{figure}
  \centering
  \includegraphics[width=.85\textwidth]{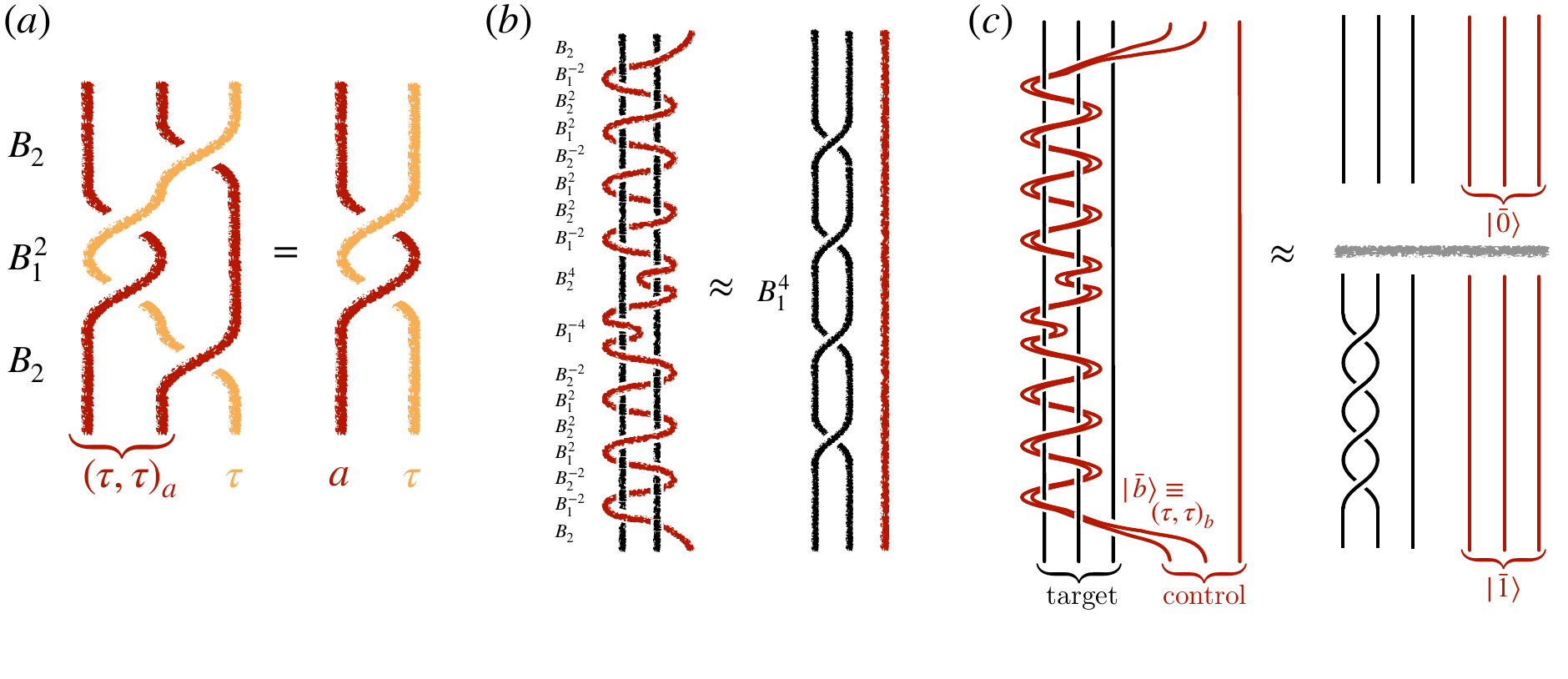}
  \caption{%
    $(a)$ The braid $B_2 B_1^2 B_2$ moves the combined object
    $(\tau_1,\tau_2)_a$ (dark) around the anyon $\tau_3$ (light). If the
    fusion outcome is the vacuum ($a=0$), the state of the system remains
    unchanged. On the other hand, if the fusion outcome is an anyon
    ($a=\tau$), the result is equivalent to $(U_{0\tau})^2$ obtained by simply
    moving an elementary anyon around the light braid. $(b)$ The braid where
    the red anyon is `woven' around the black anyons is approximately
    equivalent to $B_1^4$ with an error of $1\,\%$. $(c)$ The approximate
    braid of $(b)$ can be used to implement a two-qubit gate (a controlled
    unitary gate) on 6 anyons (the black anyons form the target and the red
    anyons the control qubit). The state of the control qubit is encoded in
    the fusion outcome of the two leftmost red braids. If the fusion outcome
    is the vacuum (the control qubit is in state $|\bar 0\rangle$), the result
    of the braid is the identity operation. If the fusion outcome is an anyon
    (the control qubit is in state $|\bar 1\rangle$), the net result is
    $U_4=(U_{0\tau})^4$ according to $(b)$.
  }\label{fig:comp_braid}
\end{figure}

To understand the idea of Ref.~\cite{bonesteel:05} regarding how to implement
two-qubit gates, we first have to appreciate another crucial property of anyon
braiding: braiding the fusion product $(\tau_1, \tau_2)_a$ as a composite
object results in the same net effect as braiding the elementary object $a$.
As a simple example, let us consider the states $|0,\tau, a ,b\rangle=
|((\tau_1,\tau_2)_a,\tau_3)_b\rangle$ as before. As seen in
Fig.~\ref{fig:comp_braid}$(a)$, the braid $B=B_2B_1^2B_2$ moves the composite
object $(\tau_1,\tau_2)_a$ around $\tau_3$ and is represented  by
$W=\rho_\tau(B)$. From Eq.~\eqref{eq:uandv}, we obtain after a straightforward
calculation the result\footnote{It follows from $V_{0\tau}
(U_{0\tau})^2V_{0\tau}  = \operatorname{diag}(1, \omega^{-2})$ and $V_{00}
(U_{00})^2V_{00} = \omega^{-4}$.}
\begin{equation}
  W_{ab} =  \begin{cases} 1 , & a=0, b=\tau,\\
    \omega^{-4},& a=\tau, b= 0, \\
    \omega^{-2},& a=\tau, b=\tau .
  \end{cases}
\end{equation}
We observe that as expected moving two anyons which fuse to the vacuum ($a=0$)
around another anyon does not change the state of the system. However, when the
fusion outcome is another anyon ($a=\tau$), the result is $(U_{0\tau})^2$.

The following procedure allows us to obtain a controlled two-qubit gate by
braiding: first, we find a `weave', i.e., a braid where only a single anyon is
moved around two static anyons that is equivalent to a braid, which only
involves the static anyons. It can be checked that the weave of the third
anyon shown in Fig.~\ref{fig:comp_braid}$(b)$ is equivalent to $B_1^4$ of the
first two anyons up to an error that is smaller than a percent. Having found
such a weave by brute force search, a controlled gate can be obtained with the
idea of the composite object explained above, see
Fig.~\ref{fig:comp_braid}$(c)$.  As before, we encode two qubits in the states
$|\bar a, \bar b \rangle = |((\tau_1,\tau_2)_a, \tau_3)_\tau \rangle \otimes
|((\tau_4,\tau_5)_b, \tau_6)_\tau \rangle $ of six Fibonacci
anyons.\footnote{In terms of our states $|f_0, f_1 , \dots\rangle$ where $f_j$
denotes the fusion outcome of all the anyons to the left, we have to add
another passive anyon, which fuses with $\tau$ to the vacuum, in between the
three anyons forming qubit $a$ and the three anyons forming qubit $b$.  As a
result, we have the encoding $|\bar a ,\bar
b\rangle=|0,\tau,a,\tau,0,\tau,b,\tau\rangle$. Without the additional fusion
outcome $0$ the value $b$ would not denote the fusion outcome of the two
anyons to the left of it.} We use $|\bar b\rangle$ (formed by anyons
$\tau_4,\tau_5,\tau_6$ in red)  as the control qubit  and $|\bar a\rangle$
(formed by anyons $\tau_1,\tau_2,\tau_3$ in black) as the target.  As the
state of the control qubit is given by the fusion outcome $b$ of
$\tau_4,\tau_5$, moving the two anyons together around the anyons of the
target qubit does nothing as long as $b=0$ (in this case, they are equivalent
to the vacuum that braids trivially). In order to obtain a controlled gate
that does not change the state of $|\bar b\rangle$, we let the pair of anyons
$\tau_4,\tau_5$ perform the weave found above.  In this case, the control
anyons (up to a small error) remain unaffected and, provided that $b=\tau$,
the operation $U_4 = (U_{0\tau})^4 = \operatorname{diag}(e^{2\pi i/5},
e^{-4\pi i/5})$ is performed on the target. As a result, the braid of
Fig.~\ref{fig:comp_braid}$(c)$ acting on two qubits entangles them by a
controlled-$U_4$ gate. Together with the universal set of single qubit gates
found before, all gates can be performed by braiding Fibonacci anyons and the
anyons are \emph{universal}\index{anyon!universal} for quantum computation.
More information on Fibonacci anyons can be found, e.g., in
Refs.~\cite{preskill,hormozi:07,nayak:08,trebst:08}. For a method to design
efficient braids that approximate a target unitary, see
Refs.~\cite{burrello:10,burrello:11}.

\section{Conclusion}

We have shown how non-Abelian anyons can be used for topological quantum
computers. The computation is performed by the wordlines of the anyons forming
knots in space-time. The outcome of the computation is the fusion result at
the last step of the computation. The computation is topologically protected in
the sense that the gates performed do not depend on the concrete trajectories
of the anyons but only on the overall topology of the braid.

The discussion so far has been concerned only with zero temperature. At any
finite temperature, there is a finite fraction $\propto \exp(-\bar M /k_B T)$
of thermal anyons present in the system ($\bar M$ is the gap protecting the
topological phase). Those anyons braid in an unwanted, uncontrollable fashion
around the anyons forming the qubit. As a result, the computation is dephased
and fails. If topological quantum computation is so fragile with respect to
temperature, why did we bother to discuss the ideas at length?  The answer to
this question has many facets.  First and most importantly, non-Abelian anyons
are an interesting new form of quantum matter.  The study of the potential
ways in which (quasi-)particles may braid, depending on the dimension of
space, is an interesting and important question of basic physics research,
even without applications to quantum computation.

Regarding the potential for quantum computation, the argument given above
simply means that we have to employ error correction at some level.
Non-Abelian anyons have inspired error correction for a long time and have
led to the concept of topological codes \cite{kitaev:03}. In particular, even
if there is no passive system that features Fibonacci anyons at zero
temperature, we can imagine keeping a system actively in a topological
state with non-Abelian anyons by constantly performing syndrome measurements
\cite{konig:10,bonesteel:12}.

Regarding the parity-protected quantum computation with Majorana zero modes;
even though the protocol strictly only works at zero temperatures, we may
expect the Majorana qubit to have a rather long lifetime at finite
temperature. Moreover, it has recently been shown that the non-Abelian nature
of the Majorana zero modes allows for dedicated error protection protocols
relying on their non-Abelian nature
\cite{terhal:12,vijay:15,plugge:16,litinski:17}. The study of systems where
the elementary excitations are particles with exotic  exchange statistics is
still a very active field with many surprises to come. Presently, there is a
big experimental push to realize Ising-anyons in the form of Majorana zero
modes in superconductor-semiconductor heterostructures, see
Ref.~\cite{prada:19}.  However, so far, no compelling evidence of particles
with non-Abelian exchange statistics has ever been observed.

I want to thank Lisa Arndt and Alex Ziesen for carefully reading the
manuscript and proposing changes that helped to considerably increase the
readability of the text.

\clearpage
\appendix
\section*{Appendix}

\begin{figure}[h]
  \centering 
  \includegraphics[width=0.95\linewidth]{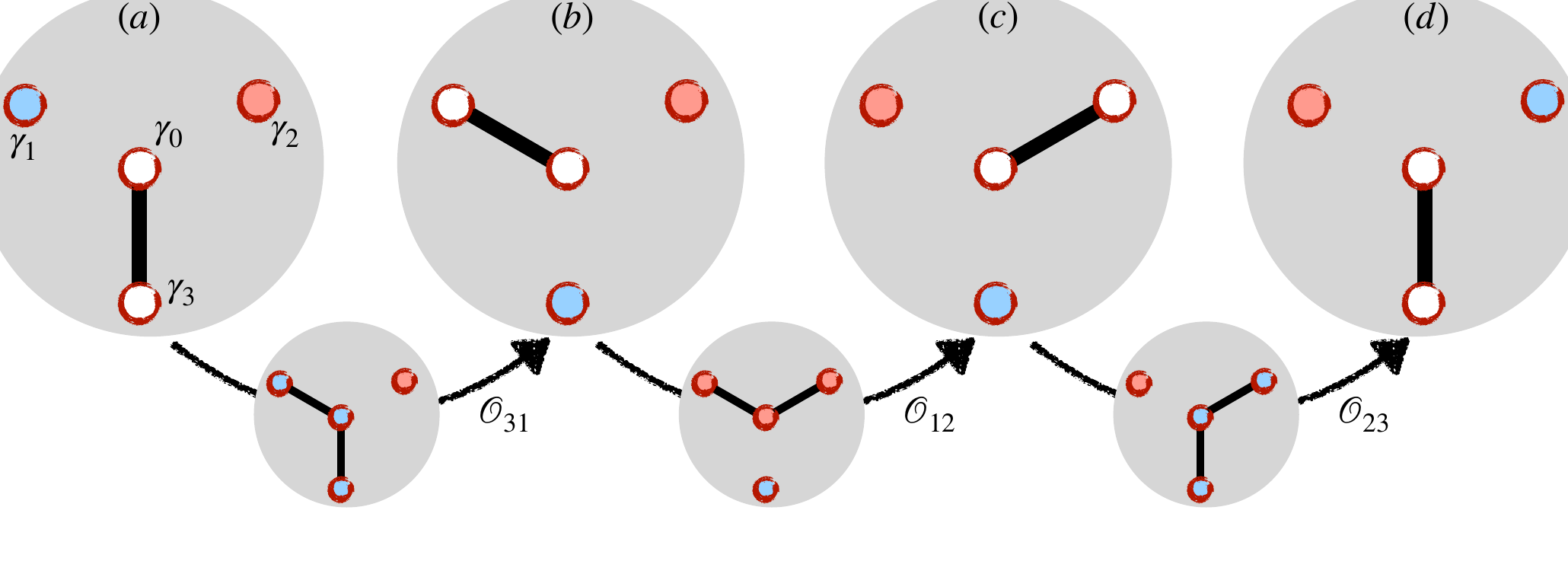}
  \caption{%
  Steps to interchange the blue and red Majorana zero modes that are initially
  located at positions 1 and 2. The first move from $(a)$ to $(b)$ moves the
  blue Majorana zero mode from position 1 to position 3. In the first step
  (small circle below), the coupling between $\gamma_0$ and $\gamma_1$ is
  increased, which leads to the delocalization of the blue Majorana zero mode
  as a superposition of $\gamma_1, \gamma_0, \gamma_3$. In the second step,
  the coupling between $\gamma_0$ and $\gamma_3$ is reduced which moves the
  blue Majorana mode to position 3. In the step from $(b)$ to $(c)$, the red
  Majorana mode is moved from position 2 to position 1. In the last step, the
  blue Majorana mode is moved from position 3 to position 2.
}\label{fig:y_braid} 
\end{figure}

\section{Braiding of Majorana zero modes}

\index{Berry phase}
In this appendix, we want to derive the unitary operation that is performed
when exchanging two Majorana zero modes. We discuss this in a system
consisting of four Majorana zero modes, which are aligned in a Y-junction
\cite{sau:11,heck:12}. We call the Majorana modes at the three ends of the
Y-junction $\gamma_1$, $\gamma_2$, and $\gamma_3$. The Majorana mode in the
middle is denoted by $\gamma_0$. The Hamiltonian of the system assumes the
form
\begin{equation}\label{eq:ham}
  H = i\sum_{j=1}^3 \epsilon_j \gamma_0\gamma_j = i \gamma_0 (\boldsymbol 
  \epsilonup \cdot \boldsymbol \gammaup)
\end{equation}
with $\epsilon_j$ as three parameters that give the coupling strength of the
two Majorana modes on each leg of the junction. Slowly tuning these couplings
in a controlled fashion makes it possible to interchange Majorana modes, and
thus we can observe the non-Abelian statistics. In particular, we would like
to study the situation in Fig.~\ref{fig:y_braid}.

We start at time $(a)$ with $\epsilon_3 = \bar\epsilon$ and
$\epsilon_1=\epsilon_2=0$. Then we increase $\epsilon_1$ to $\bar\epsilon$ and
afterwards decrease $\epsilon_3$ to 0 such that at time $(b)$ we have moved
the blue Majorana zero mode to the bottom (from position 1 to position 3). To
finish the exchange, we perform the same procedure from position 2 to position
1 and finally, from position 3 to position 2.

\subsection{Non-Abelian Berry phase}

The Berry phase is a geometric phase that arises when a parameter of the
Hamiltonian is changed slowly, and we want to observe the change in the ground
state wavefunction. In the case where the ground state is degenerate, the
change in the parameter might induce transitions between the different
degenerate states, leading to a non-Abelian Berry phase which is  generated by
the non-Abelian gauge field
\begin{equation}\label{eq:nonabelian}
  A^k_{ab}= i \langle \psi_a(\boldsymbol \epsilonup) | \partial_{\epsilon_k} 
  | \psi_b(\boldsymbol \epsilonup) \rangle;
\end{equation}
here, $\psi_a(\boldsymbol \epsilonup)$ denotes the different degenerate ground state
wavefunctions for the Hamiltonian with parameters $\boldsymbol\epsilonup$.

Changing the parameter along a  contour $\boldsymbol \epsilonup(s)$, the ground states
are transformed according to the unitary matrix
\begin{equation}\label{eq:unitary}
  U = \mathcal{P}\exp\left(i \int\!\mathbf A\cdot d \boldsymbol \epsilonup \right)
\end{equation}
where $\mathcal{P}$ denotes the path ordering.

A basis-independent way to obtain the evolution of the ground state due to
Kato is the following \cite{kato:50}: given the projector $P(\boldsymbol \epsilonup)$
onto the ground state manifold (which depends on the parameters $\boldsymbol
\epsilonup$), we define the Kato Hamiltonian
\begin{equation}\label{eq:gauge_field}
  K^k = i [ P(\boldsymbol \epsilonup), \partial_{\epsilon_k} P(\boldsymbol \epsilonup) ]
\end{equation}
with which the unitary evolution in Eq.~\eqref{eq:unitary} can be written as
\begin{equation}\label{eq:unitary_k}
  U_K = \mathcal{P} \exp\left(i \int \mathbf{K}\cdot d{\boldsymbol\epsilonup}\right).
\end{equation}
It can be shown that $U_K = U$ for any closed contour \cite{avron:89}.

\subsection[Calculation]{Calculating the non-Abelian Berry phase}

The Hamiltonian $H$ has the eigenenergies $\pm \epsilon$ ($\epsilon= |\boldsymbol
\epsilonup|$) where both of them are doubly degenerate. For the Berry phase,
we need to project onto the ground state sector on which $H=- \epsilon$. The
projector is given by
\begin{equation}\label{eq:proj}
  P(\boldsymbol\epsilonup) = \frac1{2\epsilon}(\epsilon - H) =
   \frac1{2\epsilon}[\epsilon - i\gamma_0 (\boldsymbol \epsilonup 
   \cdot \boldsymbol \gammaup)]
\end{equation}
which leads to
\begin{equation}\label{eq:proj_der}
  \partial_{\epsilon_k} P(\boldsymbol \epsilonup) = \frac{i \gamma_0}{2\epsilon^3}
  \sum_{j\neq k} (\epsilon_k \epsilon_j \gamma_j - \epsilon_j^2 \gamma_k).
\end{equation}

Now it is straightforward to calculate the Kato Hamiltonian
\begin{align}\label{eq:gauge}
  K^k &= \frac{i}{4\epsilon^4}\sum_{i} \sum_{j\neq k}
  \epsilon_i
  \Bigl(\epsilon_k \epsilon_j  
  \overbrace{[\gamma_0 \gamma_i,\gamma_0 \gamma_j]}^{2 (\delta_{ij}-1) 
  \gamma_i \gamma_j}
  - \epsilon_j^2 
  \overbrace{[\gamma_0\gamma_i,\gamma_0 \gamma_k]}^{2 (\delta_{ik}-1) 
  \gamma_i \gamma_k} 
  \Bigr) \nonumber\\
  &= \frac{i}{2\epsilon^4}\left[ \sum_{j\neq k} \epsilon_j^2
  \sum_{i\neq k} 
  \epsilon_i \gamma_i \gamma_k-
  \epsilon_k
  \sum_{i\neq j} \sum_{j\neq k}\epsilon_i \epsilon_j  \gamma_i \gamma_j 
  \right) \nonumber\\
  &=\frac{i}{2\epsilon^4} \Bigr[ (\epsilon^2 - \epsilon_k^2) 
  (\boldsymbol \epsilonup \cdot \boldsymbol \gammaup - \epsilon_k \gamma_k) \gamma_k
  +\epsilon_k^2  (\boldsymbol \epsilonup \cdot \boldsymbol \gammaup - 
  \epsilon_k \gamma_k) \gamma_k   \Bigr] \nonumber\\
  &= \frac{i}{2\epsilon^2} (\boldsymbol \epsilonup \cdot \boldsymbol \gammaup
  -\epsilon_k \gamma_k) \gamma_k.
\end{align}

The result when changing a single parameter reads
\begin{equation}\label{eq:int}
  i\int_{\epsilon_\text{start}}^{\epsilon_\text{stop}}
  \!\!\! d\epsilon_k\,K^k
  = \overbrace{(\epsilon_k \gamma_k-\boldsymbol \epsilonup \cdot 
    \boldsymbol\gammaup)}^{%
    \makebox[0pt][c]{\scriptsize\qquad\qquad%
      $=-\sum_{j\neq k} \epsilon_j \gamma_j$ (indep. of
  $\epsilon_k$)}} \gamma_k   
  \times
  \underbrace{\int_{\epsilon_\text{start}}^{\epsilon_\text{stop}}\frac{d\epsilon_k}{2
    \epsilon^2}}_{%
    \makebox[0pt][c]{\scriptsize\qquad%
      $=\frac{1}{2 \epsilon_\perp } \arctan(\epsilon_k/ \epsilon_\perp)
\big|_{\epsilon_\text{start}}^{\epsilon_\text{stop}}$}}
\end{equation}
with $\epsilon_\perp = \left( \sum_{j\neq k } \epsilon_{j}^2 \right)^{1/2}$.

A braid is constructed from elementary moves, see Fig.~\ref{fig:y_braid}.  We
exemplify the calculation of $\mathcal{O}_{31}$. Starting with
$\epsilon_3 = \bar\epsilon$ and $\epsilon_1 = \epsilon_2 = 0$, we first turn
on $\epsilon_1$ from 0 to $\bar \epsilon$ which yields ($k=1,
\epsilon_\text{start}=0, \epsilon_\text{stop}= \bar \epsilon$)
\[
U_a =  \exp\left(-\bar\epsilon \gamma_3 \gamma_1 \times  \frac1{2\bar\epsilon} 
\frac{\pi}{4}  \right)
 = \exp \left( \frac\pi8 \gamma_1 \gamma_3  \right).
\]
In the second step, we start with $\epsilon_1 = \epsilon_3= \bar \epsilon$ and
$\epsilon_2 =0$ and reduce $\epsilon_3$ down to 0. This step yields a
non-Abelian Berry phase ($k=3, \epsilon_\text{start} =\bar \epsilon,
\epsilon_\text{stop}= 0$)
\[
  U_b = \exp\left(-\bar \epsilon \gamma_1 \gamma_3 \times -\frac{1}{2\bar
    \epsilon} \frac{\pi}{4}
\right)
= \exp\left(  \frac\pi8 \gamma_1 \gamma_3 \right).
\]
So all together, we have the transformation 
\begin{equation}\label{eq:o31}
  \mathcal{O}_{31} = U_b U_a =  
  \exp\left( \frac\pi4 \gamma_1 \gamma_3 \right)
  = \frac1{\sqrt{2}} ( 1 +  \gamma_1 \gamma_3) 
\end{equation}
from $(a)$ to $(b)$ that moves the Majorana zero mode from position 1 to
position 3.  The whole operation of exchanging the particles 1 and 2 is given
by
\begin{equation}\label{eq:B}
  \rho_\gamma(B_{1}) = \mathcal{O}_{23} \mathcal{O}_{12} \mathcal{O}_{31}
  = \frac{1}{2^{3/2}} ( 1+ \gamma_3 \gamma_2) ( 1+  \gamma_2 \gamma_1)
 ( 1+ \gamma_1 \gamma_3)
 = \frac{1}{\sqrt{2}} (1 -\gamma_1 \gamma_2 )
 = e^{-\frac\pi4\gamma_1 \gamma_2}.
\end{equation}

The braiding operation $\rho(B_1)$ transforms the Majorana zero modes as
\begin{equation}
  \gamma_1 \mapsto \rho_\gamma(B_1) \gamma_1 \rho_\gamma(B_1)^\dag =  \gamma_2,
  \qquad \gamma_2 \mapsto \rho_\gamma(B_1) \gamma_2 \rho_\gamma(B_1)^\dag =
  -\gamma_1.
\end{equation}

It is a natural question to ask which of the two modes $\gamma_1$ or
$\gamma_2$ gets a minus sign. It turns out that in our setting of
Fig.~\ref{fig:y_braid}, $\gamma_2$ obtains a minus sign (it is transformed to
$-\gamma_1$) as it is only moved once in the second step ($\mathcal{O}_{12})$.
On the other hand, $\gamma_1$ is moved twice, in the first step
($\mathcal{O}_{31}$) and in the last step ($\mathcal{O}_{32}$). Indeed, in a
single step, we have
\begin{equation}
  \mathcal{O}_{31} \gamma_1  \mathcal{O}_{31}^\dag = -\gamma_3
\end{equation}
and thus $\gamma_1$ is moved to $-\gamma_3$.  As a result, we would obtain the
conjugate representation $\rho_\gamma(B_1) = e^{\frac{\pi}{4} \gamma_1
\gamma_2}$, which is often employed in the literature, if the geometry in
Fig.~\ref{fig:y_braid} would
look like a `\,\rotatebox[origin=c]{180}{Y}\,' rather than a `\,Y\,'.

\clearpage

\fancyhf{}
\fancyfoot[C]{\thepage}

\end{document}